\documentclass{IEEEtran}
\usepackage[utf8]{inputenc}
\usepackage{comment}
\usepackage{alltt}
\usepackage{todonotes}
\usepackage{tcolorbox}
\usepackage[colorlinks,urlcolor=blue,linkcolor=magenta,citecolor=red,linktocpage=true]{hyperref}
\usepackage{cite}
\usepackage{graphicx}
\usepackage{caption}
\usepackage{subcaption}

\newcommand{\SWHIDhref}[1]{\href{https://archive.softwareheritage.org/#1}{\small #1}}
\newcommand{\SWHIDex}[1]{\begin{tcolorbox}\SWHIDhref{#1}\end{tcolorbox}\noindent}

\title{Referencing Source Code Artifacts:\\ a Separate Concern in Software
  Citation}

\author{
  \IEEEauthorblockN{Roberto Di Cosmo},
  \IEEEauthorblockA{Inria and University Paris Diderot, France \\
    roberto@dicosmo.org}\\
  \and
  \IEEEauthorblockN{Morane Gruenpeter},
  \IEEEauthorblockA{University of L'Aquila and Inria, France \\
    morane@softwareheritage.org}\\
  \and
  \IEEEauthorblockN{Stefano Zacchiroli},
  \IEEEauthorblockA{University Paris Diderot and Inria, France \\
    zack@irif.fr}
}

\begin{document}

\maketitle
\begin{abstract}

  Among the entities involved in software citation, software source code
  requires special attention, due to the role it plays in ensuring scientific
  reproducibility. To reference source code we need identifiers that are not
  only unique and persistent, but also support \emph{integrity} checking
  \emph{intrinsically}. Suitable identifiers must guarantee that denoted
  objects will always stay the same, without relying on external third parties
  and administrative processes.

  We analyze the role of \emph{identifiers for digital objects} (IDOs), whose
  properties are different from, and complementary to, those of the various
  \emph{digital identifiers of objects} (DIOs) that are today popular building
  blocks of software and data citation toolchains.

  We argue that both kinds of identifiers are needed and detail the syntax,
  semantics, and practical implementation of the persistent identifiers (PIDs)
  adopted by the Software Heritage project to reference billions of software
  source code artifacts such as source code files, directories, and commits.

\end{abstract}

\begin{IEEEkeywords}
  software citation, digital preservation, reproducibility, open science,
  digital object identifier
\end{IEEEkeywords}
 \section{Introduction}
\label{sec:intro}

This article builds upon key findings on identifiers for \emph{software source
  code artifacts} from our previous work on digital
archives~\cite{swhipres2018}.  We focus on its relevance for scientific
reproducibility of experiments that rely on software, where one needs to
identify source code at various granularities, retrieve byte-identical copies
of source code artifacts described in a paper, and verify their integrity as a
prerequisite for attempting to reproduce a given result.

We recall the schema of software source code identifiers that has been
developed and deployed in the context of Software Heritage~\cite{swhcacm2018}, a
long term initiative aiming to collect, preserve, and share the entire
body of software source code and its development history. We show that these
identifiers are well-suited for addressing the need of scientific
reproducibility when source code is involved.

The requirements that emerge from this particular setting, where one needs to
handle tens of billions of different digital objects, cannot be fully satisfied
by state-of-the-art identifier schemas broadly adopted for digital publications.
Fortunately, we can leverage recursive data structures based
on cryptographic hashes, such as Merkle trees~\cite{Merkle}, to build
identifiers of digital objects that satisfy the overlapping requirements of
long-term software source code preservation and scientific reproducibility.

 \section{Referencing Source Code for Reproducibility}
\label{sec:research}

Software and software-based methods are now widely used in research fields other
than just computer science and engineering~\cite{NatureTop}.  Recognition of the
role that software plays in reproducing scientific results, and of the
insufficient availability of its source code is growing~\cite{collberg2014measuring, the-real-software-crisis}.

In the journey towards open science, and reproducible scientific results, open
and unfettered access is needed to three main kinds of research
outputs~\cite{open-science-2018-review}:
\begin{enumerate}
\item \textbf{the scientific articles};
\item \textbf{the data} used or produced in the research;
\item \textbf{the source code of the software} embodying the experiment logic.
\end{enumerate}

Preserving software source code is as essential as preserving articles and
datasets to promote both open science and reproducibility. Unfettered access to an
executable version of the software is very valuable, but \emph{source code}
must be available too: it embodies the scientific \emph{knowledge}
underlying the computational calculation.

\subsection{Software reference versus citation}

A nice presentation of the many reasons why the research community
needs to take software into account can be found in the \emph{Software Citation
  Principles}~\cite{force11citationprinciples}.

When mentioning software in publications, though, we need to distinguish
the software \emph{project}, which refers to an endeavor to develop and
maintain software artifacts, from the \emph{software artifacts} (source code, executable binaries, etc.) that are the byproducts of that
endeavor. The project is not a digital object; the resulting artifacts are.

Following~\cite{alliez:hal-02135891}, we distinguish software artifact
\emph{reference} from software project \emph{citation}, that serve different
purposes.

The main intention of a \emph{citation} is to \emph{give credit} to the authors
of a given software (as a project), whereas the function of a \emph{reference}
is to \emph{precisely identify} software artifacts, usually for reuse
purposes. The two activities are intertwined, and sometimes citing the
project without also referencing artifacts is not satisfactory. But for the
specific needs of reproducibility, referencing software artifacts is often sufficient
and may be easily done without, e.g., having to track down credit
attribution, a vastly more difficult task that is worth a research article
on its own~\cite{alliez:hal-02135891}.

In this article we focus on \emph{references} and do not address
\emph{citations} further.

\subsection{References for reuse and reproducibility}

For reproducibility, software artifact references need to be
very fine-grained, down to specific versions. And they should point to a
persistent location, available over the long term.  As noticed
in~\cite{force11citationprinciples}, the usual ways of referencing software
artifacts by just pointing to the project website or the current development
repository are largely unsatisfactory, as these locations are ephemeral.

For reproducibility we also need a system of identifiers with specific
properties that current systems do not provide.
Let's consider ACM's Artifact Review and Badging policy, described at
\url{https://www.acm.org/publications/policies/artifact-review-badging}. It
defines the following properties, in order of increasing desirability:
\begin{itemize}

\item \textbf{Repeatability:} the ability to re-run an experiment by the \emph{same}
  team using the same experimental setup---including all involved software
  artifacts. Results that are not repeatable are rarely suitable for
  publication.

\item \textbf{Replicability:} the ability to re-run an experiment by a
  \emph{different} team, reusing the described experimental setup, software
  artifacts included.

\item \textbf{Reproducibility:} the ability to re-run an experiment by a
  different team, \emph{without relying} on the experimental setup and software
  artifacts developed by the original team.

\end{itemize}

Furthermore, the following ``badges'' can be associated to software artifacts
to capture their overall quality w.r.t.~reproducibility and reuse:
\begin{itemize}

\item \textbf{Available:} software artifacts that have been made available via
  publicly-accessible long-term archives.

\item \textbf{Functional:} software artifacts that meets the specific needs of
  an experiment and are documented, consistent, complete, and exercisable [by
  third parties].

\item \textbf{Reusable:} functional (as above) software artifacts that are more
  generally useful than addressing the needs of a specific experiment.

\end{itemize}

The focus of this system is enabling researchers to reproduce and verify
results, and not giving credit to authors. Even for the most bare requirement of
\emph{repeatability}, it is necessary to have at hand a precise reference to the
software source code used for the experiment, as well as long-term archival of
the referenced artifact.

As it is now customary in modern software development, we expect that the
reference itself allows \emph{integrity checking} upon artifact retrieval,
enabling researchers that are attempting replication to rule out corruption or
tampering with the digital objects as a potential cause for
non-reproducibility.  To this end, we need a system of identifiers that depends
on \emph{no middleman}, like central registries, \emph{that could silently
  change the association between identifiers and referenced objects}.

In this paper we describe an already operational system of identifiers that
satisfies all these extra properties: \emph{high granularity, integrity, and no
  middleman}. We argue that, when used in conjunction with the long-term
archival provided by Software Heritage~\cite{swhcacm2018}, such a system
provides a suitable mechanism for referencing source code artifacts in research
articles.

 \section{Identifier systems and their properties}
\label{sec:surveyid}

This whole section recalls the key findings on identifier systems that the
authors published to the attention of the digital preservation community
in~\cite{swhipres2018}.

\subsection{Identifier systems}
\label{sec:survey-systems}

A \emph{system of identifier} is composed of a set of \emph{labels} that can be
used as references for objects and a set of \emph{mechanisms} performing some or
all of the following operations:
\begin{itemize}
\item \textbf{Generation:} create a new label
\item \textbf{Assignment:} associate a label to an object
\item \textbf{Verification:} given a label and an object, verify that they
  correspond
\item \textbf{Retrieval:} given a label, provide a means of getting a copy of
  the corresponding object
\item \textbf{Reverse lookup:} given a object, find the label that has been
  assigned to it, if any
\item \textbf{Description:} given a label, provide a means of getting metadata
  describing the corresponding object
\end{itemize}

While these mechanisms can in principle be implemented by totally independent
entities, the most common systems of identifiers conflate all these
conceptually distinct mechanisms into a single logical component
usually called \emph{resolver}.

\begin{table}[t]
\caption{Mechanism implementation in common systems of identifiers\label{table:mechanisms}}
\centering
\begin{tabular}{|l|l|l|l|l|l|l|l|}
\hline
\textbf{Mech.} / \textbf{System} & \textbf{Handle} & \textbf{DOI} & \textbf{Ark} & \textbf{PURL} & \textbf{VDOI}\\
\hline
Generation & Yes & Yes & Yes & Yes & Yes\\
Assignment & Yes & Yes & Yes & Yes & Yes\\
Verification & N.A. & N.A. & N.A. & N.A. & Yes\\
Retrieval & Yes & Yes & Yes & Yes & Yes\\
Reverse Lookup & N.A. & N.A. & N.A. & N.A. & N.A.\\
Description & Yes & Yes & Yes & N.A. & Yes\\
\hline
\end{tabular}
\end{table}

Despite the fact that the \emph{verification} mechanism is of paramount
importance in all the identification systems used in the digital landscape, we
could not find any widely used system of identifiers that provides a reliable
technical way of supporting verification, even if proposals in this sense have
been around for quite a while~\cite{Arnab2006},
see Table~\ref{table:mechanisms}. For the specific needs of scientific
reproducibility this is a relevant limitation of existing schemas.

\subsection{General properties}

The main properties of identifier systems
that are relevant for the research software reference use case are:
\begin{itemize}

\item \textbf{Uniqueness:} one object should have only one canonical
  identifier.

\item \textbf{Non ambiguity:} one identifier must denote only one object.

\item \textbf{Integrity:} in most cases, and in particular for scientific
  reproducibility, one expects the object denoted by an identifier not to be
  silently changed later on. An identifier ensures integrity if a user can
  verify that the object retrieved at any point in time is exactly the one that
  was associated to it at the beginning.

\item \textbf{Persistence:} an identifier should keep its relevant properties
  on the long term, potentially even after the object it refers to has gone
  away. This term is used in the literature to capture different ideas,
  sometimes it just covers the requirement that an identifier should not
  disappear, while in other places the concept covers even integrity and non
  ambiguity.

\item \textbf{No middleman:} to get the highest grade of resilience to
  external threats such as data loss or corruption, one should not rely on
  intermediaries for assigning identifiers in the beginning or using them later
  on (e.g., for retrieval).  The name of this property is borrowed from
  security and cryptography (see, e.g.,~\cite[Chapter~3]{applied-crypto}.

\item \textbf{Abstraction:} (opacity) early adopters of the Web started using
  URLs as persistent identifiers only to face dire consequences when it became
  evident that they are not persistent. As a consequence, recent identifier
  schemas, like DOI, Ark, or Handle, pushed the idea of identifiers that do not
  expose details that are subject to change, like the exact location of a
  resource. Similar ideas can be found in Cool URIs or PURLs. The intent is
  similar to that of Abstract Data Types in computer science, hence our
  preference for the term ``abstract'' over ``opaque''.

\item \textbf{Gratis:} many traditional identifier systems, like ISBN, charge a
  fee for each identifier; several digital systems of identifiers have similar
  provisions (e.g., DOI fees~\cite{Crossref}, but also DNS~\cite{DNS}). In the
  case of digital resources that need to be created or modified frequently, and
  especially when their amount is very large, charging a per identifier fee
  is problematic, because it creates a significant barrier to adoption and
  engenders costs that can become greater than the fixed cost of the
  infrastructure needed to maintain them.

\end{itemize}

\subsection{Discussion}\label{subsec:discussion}

Many systems of digital identifiers strive to provide \emph{uniqueness}, like
URNs, ARK and DOI~\cite{internet-id, ARK}, but they all rely on administrative
structures to ensure it~\cite{ArmsCACM2001} and none of them provides technical
guarantees. This fact leads to confusing issues like conflicting DOIs, an
official list of which is maintained at
\url{https://www.crossref.org/06members/59conflict.html}.

For \emph{non ambiguity}, most common systems of identifiers rely on
administrative care, leading to the risk that the same identifier end up
denoting different objects over time; this issue is similar to what happens for
URLs: \begin{quote} {\em there is no general guarantee that a URL which at one
    time points to a given object continues to do so}
  \begin{flushright}
    \small --- RFC 1738, Uniform Resource Locators (URL)
  \end{flushright}
\end{quote} and is quite real, which was already pointed out,
for example, by Arnab and Hutchison~\cite{Arnab2006}.

Despite the fact that the term ``persistent identifier'' is now used almost
everywhere, for most resolver-based systems \emph{persistence} is a property
that is not technically guaranteed, as one can see clearly stated for example
in~\cite{RFC3650}:

\begin{quote} \em The only operational connection between a
  handle and the entity it names is maintained within the Handle System. This
  of course does not guarantee persistence, which is a function of
  administrative care.
\end{quote}

Two of the three remaining properties, \emph{integrity} and \emph{no
  middleman}, are largely ignored (and not satisfied) by the most common
systems of identifiers:
\begin{quote} \em the DOI (or any other similar system) does not
  have any mechanism to prove that a downloaded version of the document is the
  same as the document located through the resolution
  process~\cite{Arnab2006}
\end{quote}

\emph{Abstraction} seems now a generally appraised property, while the
requirement for \emph{gratuity} seems much stronger in the librarian community
than in the scientific publishing one.

Finally, let us mention here the issues of versions and granularity. An object
may be used to create a new object that is a modification of it, and one may
want to keep track of the fact that the second one is derived from the first
one.  Some systems of identifiers allow to encode this versioning information
in the object label. Similarly, an object may be composed of several other
objects, and some systems of identifiers may want to encode in the object label
the relation of containment.

\subsection{DIOs versus IDOs}
\label{sec:dio-v-ido}

The reason why the stated requirements are so difficult to satisfy was already
contained in the following key remark by Paskin~\cite{paskin2010digital}:
\begin{quote}
  \em The term ``Digital Object Identifier'' is construed as ``digital identifier
  of an object," rather than ``identifier of a digital object'': the objects
  identified by DOI names may be of any form—digital, physical, or
  abstract---as all these forms may be necessary parts of a content management
  system. The DOI system is an abstract framework which does not specify a
  particular context of its application, but is designed with the aim of
  working over the Internet.
\end{quote}

Indeed, all the systems of identifiers that are commonplace in the scholarly
world are designed to provide digital identifiers for \emph{any kind of
  object}, including people, or organizations, that have no canonical digital
representation.  These Digital Identifiers of Objects (or \emph{DIOs}) make no
assumptions on the nature of the object they represent, and hence they inherit
all the epistemic issues of the traditional naming systems: the need of a
central authority, the complexity related to handling different manifestations
of the same conceptual object (like the PDF and the Postscript version of the
same book), and more. This fact also explains why none of the systems of
Table~\ref{table:mechanisms} supports reverse lookup.\\

\begin{figure*}[t]
  \centering
\includegraphics[width=0.9\linewidth]{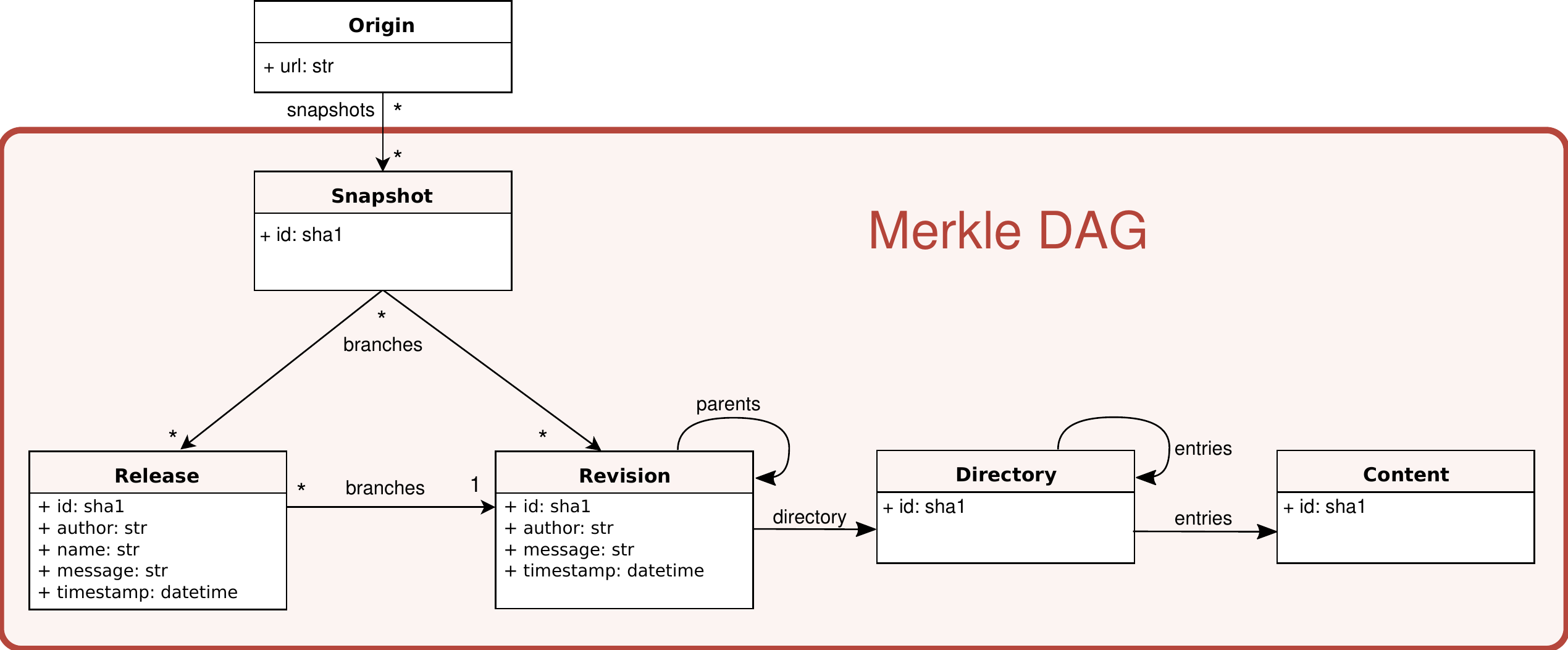}
  \caption{Topology of the Software Heritage Merkle DAG, which captures
    software source code together with its full development history.}
  \label{fig:data-model}
\end{figure*}

On the other hand, for both scientific reproducibility and software source code
archival at large, it is possible to use a system of identifiers for digital
objects (or \emph{IDO}). Such a system can be built assuming that it will only
manipulate digital objects, which means giving up the ability to attach
identifier to any kind of objects (like persons, ideas or institutions), but in
exchange all the properties that are difficult or impossible to satisfy in
traditional systems of identifiers become feasible.

The key insight is that to get all the nice properties above one must build
identifiers \emph{from the object itself}, using for example a hashing
function, and this is why we call this kind of identifiers \emph{intrinsic}.

This approach only works well if the digital object has a \emph{canonical
  representation}, on which the hashing function can be applied: it will not
be very useful for identifying documents like research articles, that can be
represented in various formats, but is perfectly suited for software artifacts.

\section{Data Model}
\label{sec:data-model}

In order to develop an identifier system for billions of source code artifacts
archived for the long-term in Software Heritage~\cite{swhcacm2018} we use a data
model based on Merkle DAG (Direct Acyclic Graph)~\cite{Merkle}. Nodes and edges
are connected using hashing functions and represent the history of software
development as captured by modern version control systems. We recall the key
concepts, and refer the reader to ~\cite{swhipres2017} for full details.

The topology of the Merkle DAG is shown in Figure~\ref{fig:data-model}. Nodes
represent: individual source code files (``content'' in the figure), source
code directories, commits (``revisions''), software releases, and entire
snapshots of the state of a given software development project. Origins are
URLs and, strictly speaking, not part of the Merkle DAG itself; but allows to
identify where source code artifacts have been encountered.

Edges between the various nodes serve different purposes depending on the type
of source/destination nodes. Edges between directories and contents simply form
on-disk source code structures. Edges between revisions (and from each revision
to the source code root directory at the time) represent the evolution of
software development over time and support both code ``forks'' and ``merges''.
Releases just annotate interesting commits at a given point in time.

In addition, various attributes (not shown in picture) are attached to nodes
and edges, depending to their types, e.g., revisions have metadata about who
did the commit and when it happened, directories attach local path names to
successors, and releases carry human-targeted labels such as ``1.0''.

\begin{figure}
  \centering
  \includegraphics[width=\linewidth]{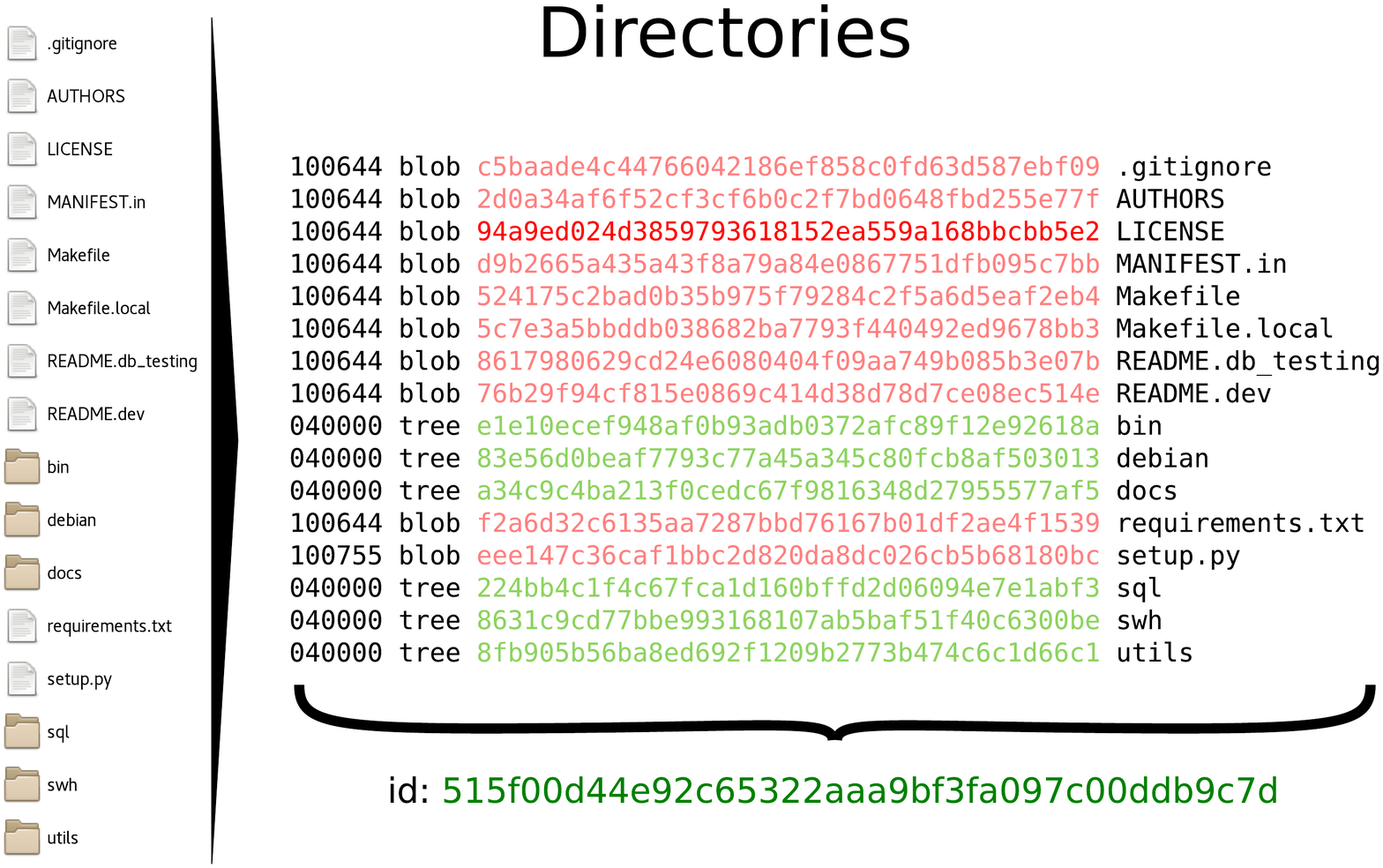}
  \caption{Intrinsic identifiers in the Merkle DAG: example of how identifiers
    for directory nodes are computed.}
  \label{fig:dir-identifier}
\end{figure}

As a consequence of the graph being a \emph{Merkle} DAG, the identifier of each
node is uniquely identified by a cryptographic checksum (SHA1 in this case).
For instance, identifiers for directory nodes are computed as shown in
Figure~\ref{fig:dir-identifier}. First, a textual serialization of the node
(known as ``manifest'') is produced; in the manifest outbound edges are
represented by the identifiers of the target nodes and metadata such as path
names are included. Then, a SHA1 checksum of the manifest is computed,
returning the desired identifier.

\begin{table*}[t]
\caption{EBNF grammar of Software Heritage persistent identifiers}
\label{tab:grammar}
\begin{alltt}
<\textbf{identifier}> ::= "swh" ":" <schema_version> ":" <object_type> ":" <object_id> ;
<schema_version> ::= "1" ;
<object_type> ::=
    "snp"  (* snapshot *)
  | "rel"  (* release *)
  | "rev"  (* revision *)
  | "dir"  (* directory *)
  | "cnt"  (* content *)
  ;
<object_id> ::= 40 * <hex_digit> ;  (* intrinsic object id, as hex-encoded SHA1 *)
<hex_digit> ::= "0" | "1" | "2" | "3" | "4" | "5" | "6" | "7" | "8" | "9"
              | "a" | "b" | "c" | "d" | "e" | "f" ;

<\textbf{identifier_with_context}> ::= <identifier> [<lines_ctxt>] [<origin_ctxt>] ;
<lines_ctxt> ::= ";" "lines" "=" <line_number> ["-" <line_number>] ;
<origin_ctxt> ::= ";" "origin" "=" <url> ;
<line_number> ::= <dec_digit> + ;
<url> ::= (* RFC 3986 compliant URLs *)  ;
\end{alltt}
\end{table*}

As a result, deduplication is built-in: if the same file appears in several
software projects, it will be represented (and hence stored) only once; its
hash will be used as unique identifier to link to its content from multiple
directories.  This process happens for directories appearing in multiple
commits, commits appearing in multiple projects, up to graph roots (i.e.,
snapshot nodes).

Since identifiers are \emph{cryptographic} checksums, node
identifiers are tamper-proof and can be used to verify the integrity of
referenced software artifacts. If, say, a file is changed during transfer or
due to media bit rot, the receiver can independently re-compute its identifier
upon reception and verify it does not match the identifier used to fetch it,
immediately realizing that the object got corrupted or has been wilfully altered.

 \section{Software Heritage Identifiers}
\label{sec:swhids}

Different experiments might need to reference and archive software source code
at different granularities: a single source code file, a tarball, a commit in a
version control system (VCS), etc. We now address the goal of how to identify
source code artifacts at all those granularities, also satisfying the
requirements of Section~\ref{sec:research}. To that end we recall in this
section the full details of Software Heritage identifiers
from~\cite{swhipres2018}.

To each source code artifact in the Software Heritage archive we associate a
\emph{persistent identifier} (PID) computed through cryptographic hashes. A PID
can point to any node in the graph described in the previous section: contents,
directories, revisions, releases, snapshots. Each PID embeds a strong
cryptographic checksum computed on the entire set of node properties and
successors, forming a Merkle structure where each node is labeled with the
identifier and provides a secure and efficient integraty mechanism.

\subsection{Syntax}

Syntactically, PIDs are generated by the EBNF grammar given in
Table~\ref{tab:grammar}.

\subsection{Semantics}

The \texttt{swh} prefix makes explicit that these identifiers are related to
Software Heritage, and the colon (\verb|:|) is used as separator between the
logical parts of identifiers. The schema version (currently \verb|1|) is the
current version of this identifier schema; future editions will use higher
version numbers, possibly breaking backward compatibility (but without breaking
the resolvability of old identifiers).

A persistent identifier points to a single object, whose type is given by
\verb|<object_type>|:
\begin{itemize}
\item \textbf{snp} identifiers points to snapshots,
\item \textbf{rel} to releases,
\item \textbf{rev} to revisions,
\item \textbf{dir} to directories,
\item \textbf{cnt} to contents.
\end{itemize}

The actual referenced object is identified by \verb|<object_id>|, which is a
hex-encoded SHA1 cryptographic checksum
computed on the content and metadata of the object itself (see
\url{https://docs.softwareheritage.org/devel/apidoc/swh.model.html} for
details).

\subsection{Examples}

\SWHIDex{swh:1:cnt:94a9ed024d3859793618152ea559a168bbcbb5e2} points to the
content of a file containing the full text of the GPL3 license.

\SWHIDex{swh:1:dir:d198bc9d7a6bcf6db04f476d29314f157507d505} points to a
directory containing the source code of the Darktable photography application
as it was at some point on 4 May 2017.

\SWHIDex{swh:1:rev:309cf2674ee7a0749978cf8265ab91a60aea0f7d} points to a commit
in the development history of Darktable, dated 16 January 2017, that added
undo/redo supports for masks.

\SWHIDex{swh:1:rel:22ece559cc7cc2364edc5e5593d63ae8bd229f9f} points to
Darktable release 2.3.0, dated 24 December 2016.

\SWHIDex{swh:1:snp:c7c108084bc0bf3d81436bf980b46e98bd338453} points to a
snapshot of the entire Darktable Git repository taken on 4 May 2017 from
GitHub.

\subsection{Contextual information}
\label{sec:context}

It is often useful to complement persistent identifiers with contextual
information about the object's setting.

One can do so with Software Heritage identifiers, using the semicolon
(\verb|;|) in the grammar as separator between PID and contextual information.
Each piece of contextual information is specified as a key/value pair, using
the equal sign (\verb|=|) as a separator. The following pieces are supported:
\begin{itemize}

\item \textbf{Software origin:} a URL where a given object has been observed in
  the wild.

\item \textbf{Line numbers:} a line number or range, pointing \emph{within} the
  given object.

\end{itemize}

For example, the following identifier
\SWHIDex{swh:1:dir:c6f07c2173a458d098de45d4c459a8f1916d900f;\\
  origin=https://github.com/id-Software/Quake-III-Arena}points to the source code root directory of the computer game Quake III Arena
with the origin URL where it was found; while
\SWHIDex{swh:1:cnt:41ddb23118f92d7218099a5e7a990cf58f1d07fa;\\lines=64-72}
points to a comment segment with the warning ``NOLI SE TANGERE'' in a file in
the Apollo-11 source code.

\subsection{Resolution}
\label{ssec:resolution}

Persistent identifiers are not directly browsable URLs, but they can be
resolved in various ways.  Any identifier can be given to the Software Heritage
Web user interface via the URL pattern
\url{https://archive.softwareheritage.org/<identifier>} to reach the referenced
object. Both in-browser and programmatic use via a dedicated REST API
endpoint is available.

The following third-party resolvers also support resolution of Software
Heritage persistent identifiers:
\begin{itemize}

\item \textbf{Identifiers.org}

\item \textbf{Name-to-Thing (N2T)}

\end{itemize}

\subsection{Verification}

Software Heritage identifiers can be generated and verified independently by
anyone using the open source \texttt{swh-identify} tool, developed by Software
Heritage and distributed via PyPI as \texttt{swh.model} (Software
Heritage identifier \SWHIDhref{swh:1:rev:6cab1cc81118877e2105c32b08653509475f3eaa;\\
  origin=https://pypi.org/project/swh.model/}).
 \section{Validation}
\label{sec:validation}

\begin{figure}
  \begin{subfigure}{\linewidth}
    \centering
    \includegraphics[width=0.9\linewidth]{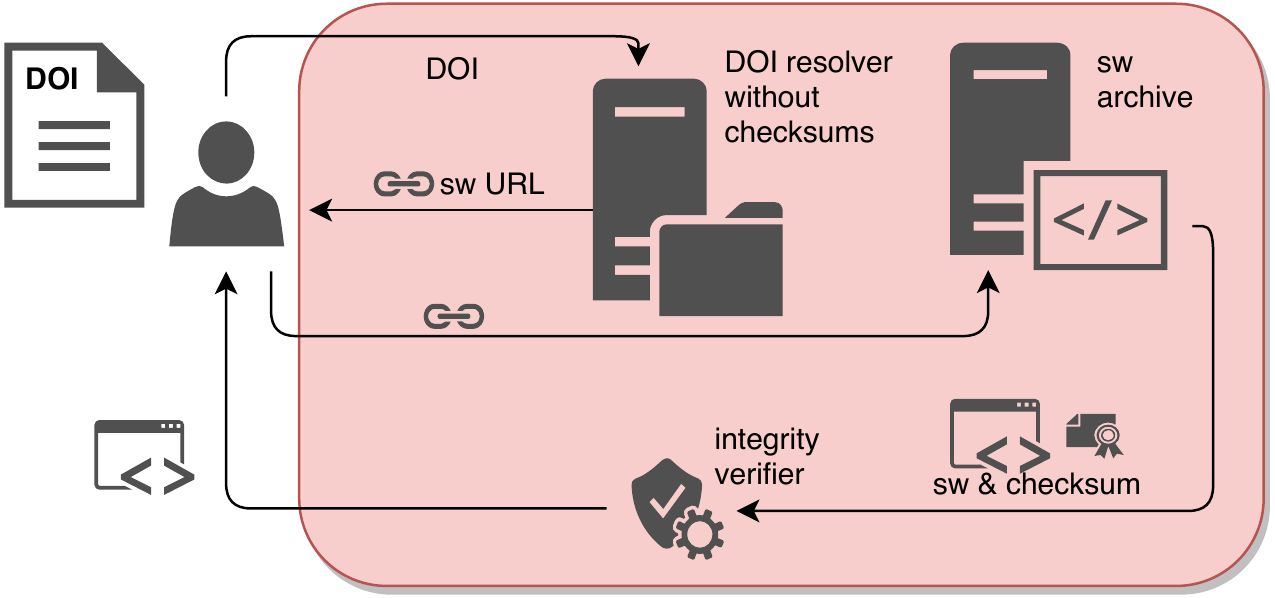}
    \caption{using DOI resolver without checksum metadata}
    \label{fig:trusted-doi-no-checksum}
  \end{subfigure}
  \vspace{2ex}
  \begin{subfigure}{\linewidth}
    \centering
    \includegraphics[width=0.9\linewidth]{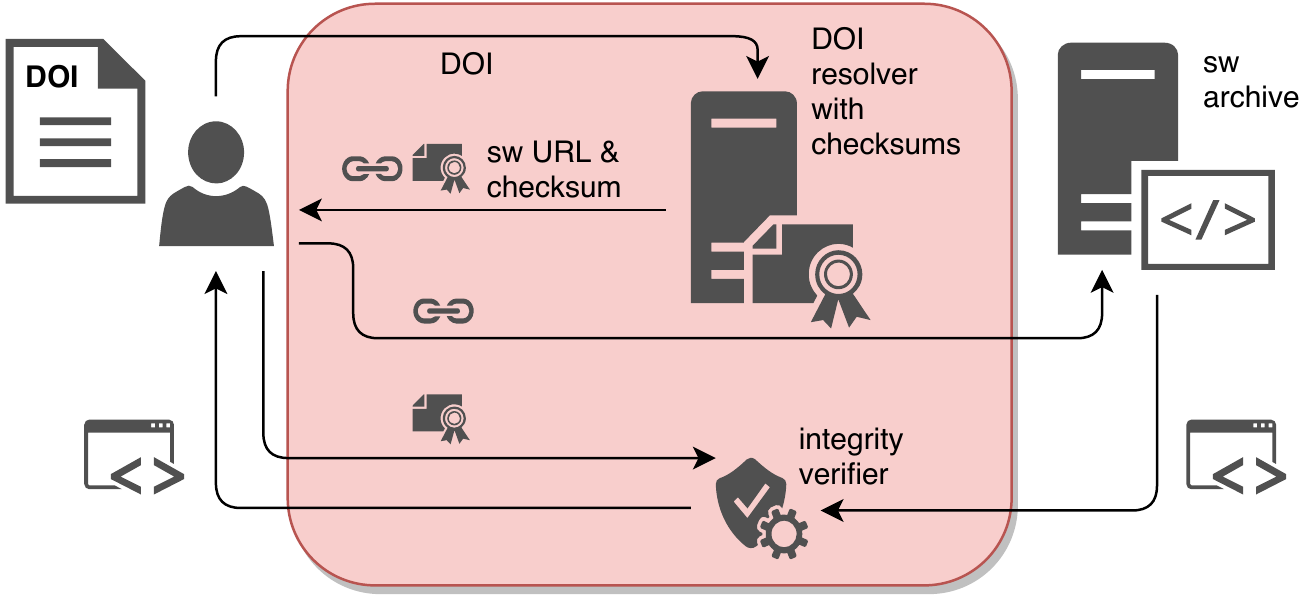}
    \caption{using DOI resolver with checksum metadata}
    \label{fig:trusted-doi-w-checksum}
  \end{subfigure}
  \vspace{2ex}
  \begin{subfigure}{\linewidth}
    \centering
    \includegraphics[width=0.55\linewidth]{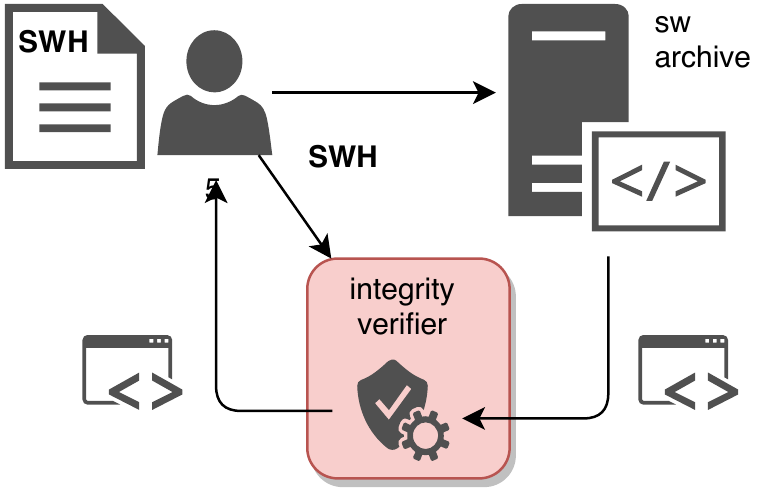}
    \caption{using Software Heritage identifiers}
    \label{fig:trusted-swh}
  \end{subfigure}
  \caption{Trusted third parties (shown as rounded red boxes) for software
    artifact retrieval and verification in three different scenarios.}
  \label{fig:trusted}
\end{figure}

We recall in this section the findings of the self-assessment exercise against
the properties discussed in Section~\ref{sec:surveyid} that was performed
in~\cite[Section~6]{swhipres2018}: we refer the interested reader to it for
more details on hash collisions.

\textbf{Uniqueness:} identifiers are computed using a cryptographic hash. By
construction and due to the Birthday paradox the chances of giving the same
identifier to different objects are negligible.

\textbf{Non ambiguity:} at each granularity level, each identifier is
designating only one object, without ambiguity.

\textbf{Persistence:} the Software Heritage archive guarantees that nothing
will be deleted intentionally and will undertake the task of perpetually
maintain old version of the identifier schema, even when new versions of it
will be released.

\textbf{Integrity:} using a cryptographic hash as identifier ensures that
modifications to the denoted object, however minimal, would yield a different
identifier with an extremely high probability. Users can recompute identifiers
on retrieved objects and verify they match.

\begin{figure*}
  \begin{subfigure}[T]{0.59\textwidth}
    \centering
    \includegraphics[width=\linewidth]{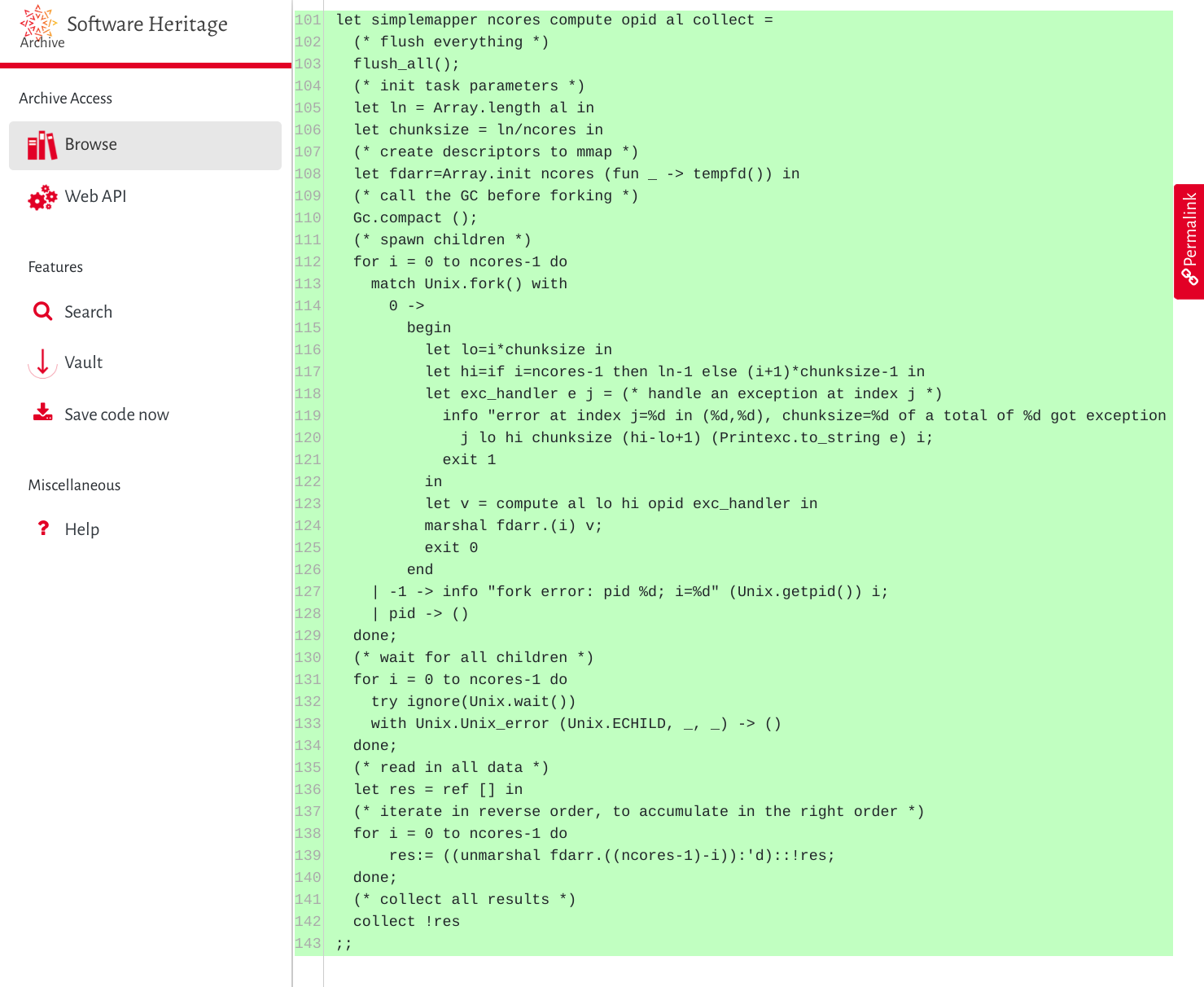}
    \caption{as archived in Software Heritage}\label{fig:parmapswh}
  \end{subfigure}
  \begin{subfigure}[T]{0.39\textwidth}
    \centering
    \includegraphics[width=\linewidth]{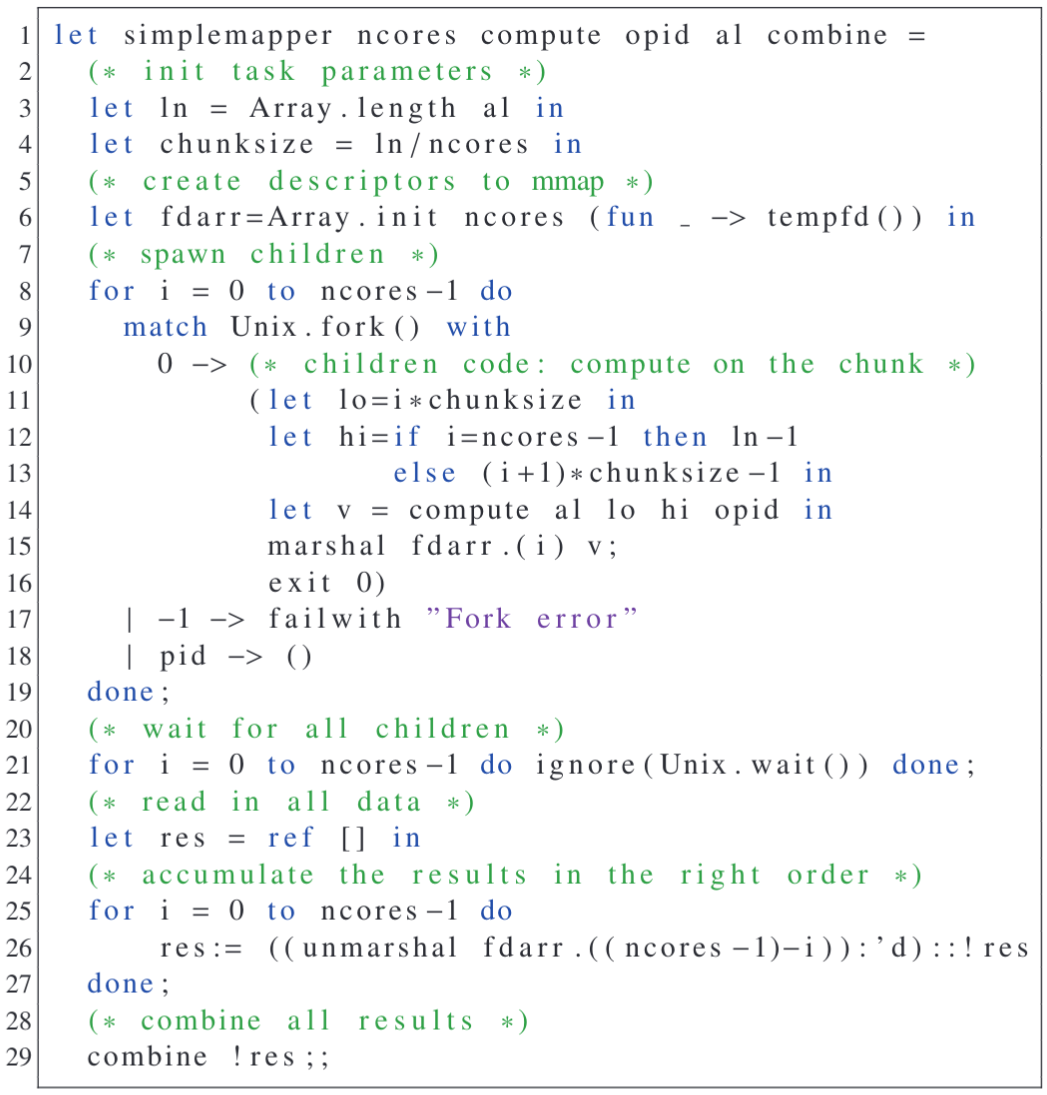}
    \caption{as presented in the original
      article~\cite{Parmap2012}}\label{fig:parmappaper}
  \end{subfigure}
  \caption{Code fragment from the published article compared to the content
    in the Software Heritage archive}
  \label{fig:parmap}
\end{figure*}

\textbf{No middleman:} the link between an object to its identifier does not
depend on the resolution of an online service. These identifiers can be used
and verified outside the system that creates and maintains them.

Figure~\ref{fig:trusted} compares the proposed approach with the
state-of-the-art in terms of third parties and communication channels that
should be trusted to verify the integrity of source code artifacts.  In the
common case of DOI resolvers that do not include artifact checksums as part of
metadata (Figure~\ref{fig:trusted-doi-no-checksum}), one has to trust the
entire toolchain. Storing artifact checksums as part of DOI metadata
(Figure~\ref{fig:trusted-doi-w-checksum}) is a significant improvement, in
which the artifact archive is no longer trusted: tampering there (or in the
communication with it) can be detected; DOI intermediaries still have to be
trusted though. The proposed approach (Figure~\ref{fig:trusted-swh}) minimizes
the trusted parties and channels: only a reliable checksum verifiers is
needed---and several exist already.

\textbf{Abstraction:} the proposed identifier schema does not expose any piece
of information that is subject to change over time.

\textbf{Gratis:} the proposed identifiers are intrinsic, meaning they can be
independently computed by anyone, using freely available software, incurring no
costs for identifier creation or attribution. By construction the obtained
identifiers will be the same everywhere, allowing cross-referencing.

\medskip Hence, we argue that the Software Heritage identifier schema provides
a systems of identifiers for digital objects (IDO) that satisfies the stated
requirements for scientific reproducibility and long-term source code
preservation.

Note that the optional contextual information of Section~\ref{sec:context} are
not strictly needed for reproducibility, but it is convenient to store extra
information, like the location from where the archived source code has been
obtained, to allow tracking future evolution of referenced software artifacts.

\section{Showcase}
\label{sec:showcase}

Software Heritage supports exact referencing of source code artifacts in two
unique ways: on the one hand, it provides a \emph{universal archive} that
stores the source code, and its full development history; on the other hand,
it uses \emph{the same intrinsic identifiers} for all its 10 billion contents, no
matter where the source code comes from, or the version control system used
to develop it.

We now look at a real world example of how this can significantly improve the
workflow of referencing source code in research articles for the purpose of
scientific reproducibility.

In 2011, Marco Danelutto and the first author started work on Parmap, a
minimalist OCaml library that implements a map-reduce framework for multicore
architectures in a concise and elegant way. The software project was developed
using git on the Gitorious forge, and described in the paper \emph{A ``minimal
  disruption'' skeleton experiment: seamless map \& reduce embedding in
  OCaml}~\cite{Parmap2012}. In order to make the code available to all and
facilitate reuse the article, published in June 2012, pointed to the open
source release of Parmap linking to~\url{https://gitorious.org/parmap}.

Alas, Gitorious was closed down in June
2015, and that URL is now broken. Luckily, Software Heritage has archived
\texttt{parmap} along with all repositories from Gitorious: that repository,
with all its development history, can now be recovered. The same can be done
for all other legacy articles referencing code on that lost platform. The
\emph{universal archive} functionality is essential, as the code can be
salvaged without requiring proactive actions by researchers.

The unique identifiers provided by Software Heritage allow to go much further,
and enable \emph{precise traceability of code versions and fragments therein}.
In Figure 1 of the Parmap article, the authors show the core part of the code
implementing the parallel functionality, consisting of 29 lines. In 2012, we
had no way to reference these exact lines in the version of the code associated
to the published article.

Today, using the proposed identifiers, the same code fragment can be precisely
identified as:
\SWHIDex{swh:1:cnt:d5214ff9562a1fe78db51944506ba48c20de3379;\\
  origin=https://gitorious.org/parmap/parmap.git;\\
  lines=101-143}

Figure~\ref{fig:parmapswh} shows side-by-side the code as archived by Software
Heritage and as shown in the paper, allowing to notice that the code in the
article was slightly simplified w.r.t.~the actual implementation. Today, a
corresponding clickable link could be easily added in the caption of the figure
(see page 5 of an \href{http://www.dicosmo.org/share/parmap_swh.pdf}{updated
  version} of the original article).

\begin{figure}[t]
  \centering
  \includegraphics[width=\linewidth]{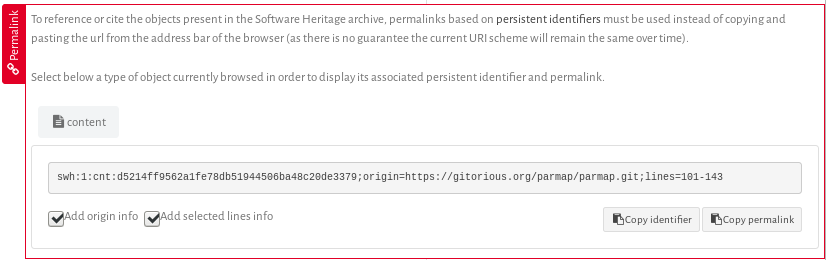}
  \caption{Obtaining a Software Heritage identifier using the permalink box on
    the archive Web user interface}
  \label{fig:permalink}
\end{figure}

Software Heritage identifiers are easy to obtain using the permalink box on the
archive's interface, as shown in Figure~\ref{fig:permalink}.

The authors should also reference in their article the exact \emph{version} of
the software project containing the code, which is also possible using a
Software Heritage revision identifier:
\SWHIDex{swh:1:rev:0064fbd0ad69de205ea6ec6999f3d3895e9442c2;\\
  origin=https://gitorious.org/parmap/parmap.git }

 \section{Conclusion}
\label{sec:future}

Software is an important product of research, and needs to be properly
mentioned in research articles, both to give academic credit to the persons
involved and to support reproducibility of research. We consider that these two
concerns are both important, but separate: while \emph{citations} are essential
for giving credit, \emph{references} are sufficient for reproducibility.

In this article, we have focused on the key properties that \emph{references}
need to satisfy in the context of scientific reproducibility, some of which
traditional digital identifiers of an object (DIOs) do not enjoy, in particular
the ability to independently verify object integrity.

Cryptographic hashes widely used in software development can be used as
identifiers of digital objects (IDOs) that satisfy all the key requisites, and
lie at the core of the Software Heritage identifier schema that is used in
production to identify over 10 billion different objects in the project
archive.

We look forward to wider adoption of these IDOs in the research community for
software artefacts, and all digital objects that have a canonical representation.


\end{document}